\documentclass{emulateapj}
\usepackage{apjfonts, natbib}

\newcommand{\be}{\begin{equation}}
\newcommand{\ee}{\end{equation}}

\newcommand{\ba}{\begin{eqnarray}}
\newcommand{\ea}{\end{eqnarray}}

\shorttitle{Improved parameters from QUaD}
\shortauthors{QUaD collaboration - S. Gupta et al.}

\begin{document}

\slugcomment{Accepted to Apj}

\title{Parameter Estimation from Improved Measurements of the Cosmic Microwave Background from QUaD}
\author{
 QUaD collaboration -- 
  S.\,Gupta\altaffilmark{1},
  P.\,Ade\altaffilmark{1},
  J.\,Bock\altaffilmark{2,3},
  M.\,Bowden\altaffilmark{1,4},
  M.\,L.\,Brown\altaffilmark{5},
  G.\,Cahill\altaffilmark{6},
  P.\,G.\,Castro\altaffilmark{7,8},
  S.\,Church\altaffilmark{4},
  T.\,Culverhouse\altaffilmark{9},
  R.\,B.\,Friedman\altaffilmark{9},
  K.\,Ganga\altaffilmark{10},
  W.\,K.\,Gear\altaffilmark{1},
  J.\,Hinderks\altaffilmark{4,11},
  J.\,Kovac\altaffilmark{3},
  A.\,E.\,Lange\altaffilmark{3},
  E.\,Leitch\altaffilmark{2,3},
  S.\,J.\,Melhuish\altaffilmark{12},
  Y.\,Memari\altaffilmark{8},
  J.\,A.\,Murphy\altaffilmark{6},
  A.\,Orlando\altaffilmark{1,3}
  C.\,O'\,Sullivan\altaffilmark{6},
  L.\,Piccirillo\altaffilmark{12},
  C.\,Pryke\altaffilmark{9},
  N.\,Rajguru\altaffilmark{1,13},
  B.\,Rusholme\altaffilmark{4,14},
  R.\,Schwarz\altaffilmark{9},
  A.\,N.\,Taylor\altaffilmark{8},
  K.\,L.\,Thompson\altaffilmark{4},
  A.\,H.\,Turner\altaffilmark{1},
  E.\,Y.\,S.\,Wu\altaffilmark{4}
  and
  M.\,Zemcov\altaffilmark{1,2,3} 
}

\altaffiltext{1}{School of Physics and Astronomy, Cardiff University,
  Queen's Buildings, The Parade, Cardiff CF24 3AA, UK.}
\altaffiltext{2}{Jet Propulsion Laboratory, 4800 Oak Grove Dr.,
  Pasadena, CA 91109, USA.}
\altaffiltext{3}{California Institute of Technology, Pasadena, CA
  91125, USA.}
\altaffiltext{4}{Kavli Institute for Particle Astrophysics and
  Cosmology and Department of Physics, Stanford University,
  382 Via Pueblo Mall, Stanford, CA 94305, USA.}
\altaffiltext{5}{Astrophysics Group, Cavendish Laboratory,
  University of Cambridge, J.J. Thomson Avenue, Cambridge CB3 OHE,
  UK.}
\altaffiltext{6}{Department of Experimental Physics,
  National University of Ireland Maynooth, Maynooth, Co. Kildare,
  Ireland.}
\altaffiltext{7}{{\em Current address}: CENTRA, Departamento de F\'{\i}sica,
  Edif\'{\i}cio Ci\^{e}ncia, Piso 4, Instituto Superior T\'ecnico -
  IST, Universidade T\'ecnica de Lisboa, Av. Rovisco Pais 1, 1049-001
  Lisboa, Portugal.}
\altaffiltext{8}{Institute for Astronomy, University of Edinburgh,
  Royal Observatory, Blackford Hill, Edinburgh EH9 3HJ, UK.}
\altaffiltext{9}{Kavli Institute for Cosmological Physics,
  Department of Astronomy \& Astrophysics, Enrico Fermi Institute, 
  University of Chicago, 5640 South Ellis Avenue, Chicago, IL 60637, USA.}
\altaffiltext{10}{APC/Universit\'e Paris 7 -- Denis Diderot/CNRS,
  B\^atiment Condorcet, 10, rue Alice Domon et L\'eonie Duquet, 75205
  Paris Cedex 13, France.}
\altaffiltext{11}{{\em Current address}: NASA Goddard Space Flight
  Center, 8800 Greenbelt Road, Greenbelt, MD 20771, USA.}
\altaffiltext{12}{School of Physics and Astronomy, University of
  Manchester, Manchester M13 9PL, UK.}
\altaffiltext{13}{{\em Current address}: Department of Physics and Astronomy, University College London, Gower Street, London WC1E 6BT, UK.}
\altaffiltext{14}{{\em Current address}:
  Infrared Processing and Analysis Center,
  California Institute of Technology, Pasadena, CA 91125, USA.}


\begin{abstract}
We evaluate the contribution of cosmic microwave background (CMB) polarization spectra to cosmological parameter constraints. 
We produce cosmological parameters using high-quality CMB polarization data from the ground-based QUaD experiment and demonstrate for the majority of parameters that there is significant improvement on the constraints obtained from satellite CMB polarization data. 
We split a multi-experiment CMB data set into temperature and polarization subsets and show that the best-fit confidence regions for the $\Lambda$CDM six-parameter cosmological model are consistent with each other, and that polarization data reduces the confidence regions on all parameters. 
We provide the best limits on parameters from QUaD EE/BB polarization data and we find best-fit parameters from the multi-experiment CMB data set using the optimal pivot scale of $k_p=0.013$~Mpc$^{-1}$ to be \{$h^2\Omega_c$, $h^2\Omega_b$, $H_0$, $A_s$, $n_s$, $\tau$\}=\{0.113, 0.0224, 70.6, 2.29$\times 10^{-9}$, 0.960, 0.086\}.
\end{abstract}



\section{Introduction}
\label{sec:intro}

 The cosmic microwave background radiation contains fluctuations in temperature and polarization which have specific spectral features that record the evolution and constituent properties of the universe. The radiation is predicted to be polarized at the 10\% level due to Thomson scattering in the presence of velocity inhomogeneities in the photon-baryon fluid at last scattering.

 The standard cosmological model predicts acoustic peaks in CMB intensity and polarization spectra. Polarized CMB radiation can be decomposed into even-parity E-modes which are generated by scalar and tensor perturbations, and odd-parity B-modes which are generated by gravitational waves and gravitational lensing effects. In this paper we examine the cosmological implications polarization spectra including the spectra produced by the 2009 improved analysis of QUaD second and third season observations presented in \cite{Michael09}. QUaD improves on the detections of E modes made by the DASI \citep{kovac02}, CAPMAP \citep{Hedman}, Boomerang \citep{jones06}, Wilkinson Microwave Anisotropy Probe (WMAP) \citep{dunkleypol}, and CBI \citep{readhead04} experiments, adding accurate polarization data at small angular scales.

\begin{deluxetable*}{l c c c}[h]
  \tablewidth{10cm}
  \tablecaption{
	Baseline and Derived Parameters and Flat Priors on Baseline Parameter Set 
\label{tab:parampriors}}
\tablehead{\colhead{Description}& \colhead{Parameter} & \colhead{Prior} &} 
    \startdata
    Baryon density &$\Omega_bh^2$ & $0.001 - 0.999 $  &\\
    Cold dark matter density &$\Omega_ch^2$ & $0.001 - 0.999 $  &\\
    Acoustic peak scale &$\theta$ & $ 0.3 - 12 $  &\\
    Scalar fluctuation amplitude & ln$(10^{10}A_s)$ & $ 2.7 - 4 $  &\\
    Scalar fluctuation index &$n_s$ & $ 0.01 - 2 $  &\\
    Optical depth &$\tau$ & $ 0.01 - 0.8 $  &\\
    Age & Age (GYr) &$ 10 - 20 $&\\
    Dark energy density & $\Omega_\Lambda$ &... &\\
    Matter density & $\Omega_m$ &...&\\
    Reionization depth & $z_{re}$&...&\\
    Hubble constant & $H_0$ & 40 - 100 &\\
    Linear mass perturbation & $\sigma_8$&...&\\
   \enddata
\end{deluxetable*}

\begin{figure}[h!tp]
\resizebox{\columnwidth}{!}{\includegraphics[width=52mm]{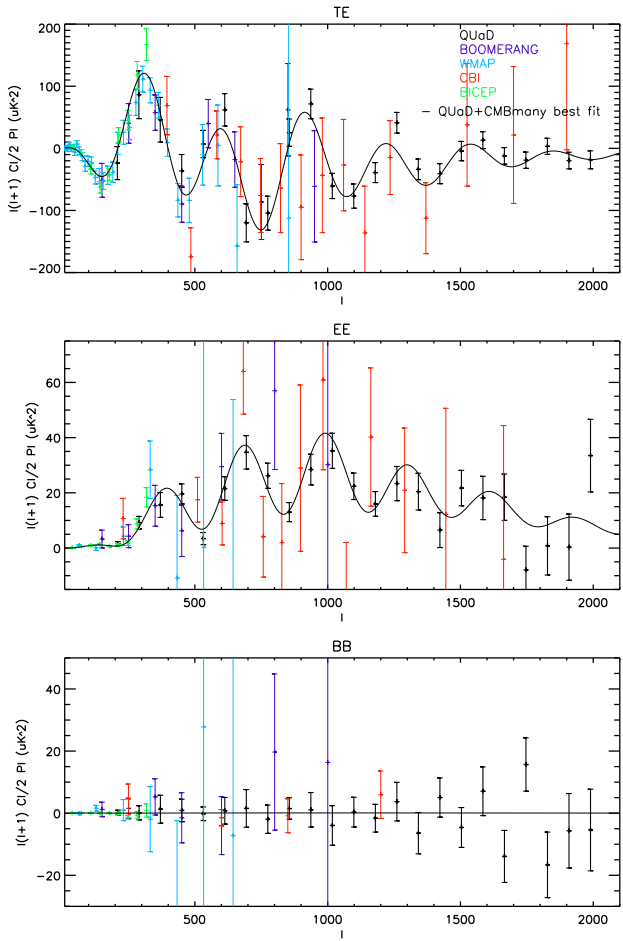}}
\caption{QUaD polarization spectra (black) plotted together with WMAP, Boomerang, CBI and BICEP polarization spectra. The black lines are the QUaD+CMBmany theoretical best-fit spectra.}
\label{fig:spec_pol}
\end{figure}

We presented design and optics reports describing the QUaD\footnote[1]{QUaD stands for ``QUEST and DASI''. In turn, QUEST is ``Q \& U Extragalactic 
Survey Telescope'' and DASI stands for ``Degree Angular Scale Interferometer''. } experiment in \cite{Hinderks09} and \cite{osullivan08}. The first season of QUaD results appeared in \cite{ade07}. We first reported measurements of temperature and polarization from the second and third season QUaD data in \cite{pryke09} and cosmological parameter analysis of the data was carried out in \cite{Castro09}. The second and third season data sets from the QUaD experiment have been reanalyzed using a ground template removal technique rather than field-differencing to remove ground contamination. A description of the method and maps and spectra from two parallel and independent pipelines can be found in \cite{Michael09}. The re-analysis has effectively doubled the size of the QUaD field, increasing the precision of the CMB spectra measurements by $\sim 30\%$  and constraining the amplitude of the lensing B-modes to <$0.57\mu K^2$ at 95\% confidence. In \cite{Michael09} we also presented a cosmological parameter analysis which gauged the effect of using the improved spectra from QUaD in combination with WMAP and ACBAR data, demonstrating that QUaD data adds significant power for constraining cosmological models which include tensors and running of the spectral index.  

In this paper we carry out cosmological parameter estimation in which we explore separately the strengths of CMB temperature data and CMB polarization data including the 2 year data release by BICEP \citep{BICEP}, highlighting the contributions of each to a multi-experiment analysis. 

\section{Data and Method}
\label{sec:method}

 The QUaD data set consists of 143 selected days of data measured during the Austral winters of 2006 and 2007, at 100 GHz and 150 GHz, in a region of approximately 100 deg$^2$, and has produced polarization spectra of unprecedented quality \citep{pryke09, Michael09}. The QUaD experiment is the first experiment with the sensitivity to detect multiple acoustic oscillations in the E-mode spectrum and TE-spectrum up to $\ell$=2000 in addition to providing the lowest upper-limits on B-mode detection \citep{Michael09}. The spectra we used for cosmological data analysis, TT, TE, EE and BB, are an optimally weighted combination of three sets of spectra: 100 GHz, 150 GHz and 100GHz-150GHz cross, each with 23 band power values in the range $200 < \ell < 2000$. 

Our CosmoMC-based parameter analysis is broadly similar to the Monte-Carlo Markov chain analysis of \cite{Castro09}, hereafter Parameter Paper 1. A single, constant covariance matrix was estimated from simulations within the QUaD pipeline. The covariance matrix is populated only in the diagonal, 1st off-diagonal and 2nd off-diagonal terms of each of the sub-blocks of the spectra. TT-TE and TE-EE have no 2nd off-diagonal terms and TT-EE and EE-BB have covariance terms up to the 12th and 4th band powers respectively. Beyond these regions the covariance matrix would be dominated by noise from the numerical simulations. We also generated offset-lognormal x-factors from noise-only simulations. This enables us to model the offset-lognormal likelihood for our parameter estimation as suggested by \cite{BJK2000}. 

We make use of templates of the Sunyaev-Zel'dovich (SZ) amplitude derived from \cite{SZ} which model a frequency-dependent contribution to the temperature power spectrum from the thermal SZ effect\footnote[2]{SZ templates are available at http://lambda.gsfc.nasa.gov.}. We apply a cut of spectral power above $\ell$=2000 to avoid the effects of residual point-source contamination, and analytically marginalize over the SZ amplitude\footnote[3]{Our analysis in Parameter Paper 1 did not include marginalization over an SZ amplitude or offset-lognormal factors.}. The justification of a fit to SZ parameters  on scales below $\ell$=2000 has not been confirmed by the high multipole temperature data obtained by QUaD \citep{Friedman09}. 

We estimate likelihoods from our CMB distribution using the publicly available CosmoMC \citep{lewis02} Monte Carlo Markov chain algorithm. The theoretical CMB model at every stage is obtained from the publicly available CAMB Boltzmann code \citep{lewis00}. The choice of baseline parameter set, and the shape of the priors on parameters fitted concurrently will impact on the one-dimensional (1D) marginalized parameters calculated using Monte Carlo Markov chain. For the standard $\Lambda$CDM six-parameter cosmological model, we limit this effect by using the {\it baseline} parameter combination comprising the baryon density, cold dark matter density, acoustic peak scale which records the expansion history of the universe, scalar spectral amplitude, scalar spectral index and optical depth: $\{h^2 \Omega_b, h^2\Omega_c, \theta,$ ln$(10^{10}A_s), n_s, \tau\}$, the use of ln$(10^{10}A_s)$ also considerably increases the rate of convergence; $H_0=100h$km s$^{-1}$ Mpc$^{-1}$ is the Hubble constant. This is the default parameter set of CosmoMC \citep{lewis02}. 

When presenting two-dimensional (2D) plots of likelihood space evaluated from a CMB spectrum that does not have information in the angular range below the first acoustic peak, we use the combined $A_s \exp{(-2\tau)}$ parameter, a measure of the overall amplitude of the fluctuations in the spectra optimally sampled in the analysis using the baseline parameter set. The pivot scale used to evaluate the amplitude and spectral index is $k_p=0.05$~Mpc$^{-1}$ in the case where we use QUaD data exclusively. The pivot scale used for multi-experiment CMB spectra analysis is $k_p=0.013$~Mpc$^{-1}$ in common with our cosmological parameter results of \cite{Michael09}. We impose flat priors on the baseline parameter set (Table \ref{tab:parampriors}), make the assumption that the universe is flat and include in all analyses the effects of weak gravitational lensing.

In addition we make use of the CosmoMC utility to analytically marginalize over ``nuisance'' parameters, which is implemented for beam and calibration uncertainties \citep{bridle02}. The effective beam size for the combined QUaD spectra is 4.1 arcmin. The beam error is a function dominated by the sidelobes and varies with scale. The calibration uncertainty is 6.8\% in power.

We provide parameter constraints on these baseline parameters and additionally include plots and constraints in the more traditional format of  $\{h^2 \Omega_b,  h^2\Omega_c, H_0, A_s, n_s, \tau\}$.

We present 68\% and 95\% confidence regions for cosmological parameters and 2D and 1D marginalized plots using our independent statistics code on 10 or more Monte-Carlo Markov chains with at least 100,000 converged steps. Tests of convergence were carried out using the Gelman-Rubin statistic \citep{gelman92}. We also made use of the Getdist statistics package, which is bundled with CosmoMC, as a consistency check of our results.

 We carried out a parameter fit from the products of both QUaD data analysis pipelines to confirm consistency. 

\subsection{Other Data}
\label{sec:other}

We combine QUaD with the following CMB data sets: ACBAR \citep{reichardt08}, including the 2\% beam error on the 5 arcmin beam, and calibration error 4.6\% in power; CBI \citep{Sievers09} with a stated calibration error of 2.6\% in power; BICEP \citep{BICEP}, 5.6\%  calibration error and beam error varying with scale; the WMAP 5 year release \citep{nolta08} using the publicly available WMAP 5 year likelihood software\footnote[4]{WMAP likelihood software is available at http://lambda.gsfc.nasa.gov.}. We describe the set of CMB data above, excluding QUaD, as {\it CMBmany}. In Figure \ref{fig:spec_pol}, QUaD polarization spectra are plotted together with WMAP, CBI, Boomerang\footnote[5]{QUaD shares its entire survey region with Boomerang, therefore we do not include Boomerang spectra in CMBmany.} and BICEP polarization data.

\section{Results}
\label{sec:results}


\begin{deluxetable*}{l c c c c c}[p!t]
  \tablewidth{12cm}
  \label{sec:maintable}
  \tablecaption{
	Cosmological Parameter Constraints Using QUaD Data  
\label{tab:paramsQUaD}}
\tablehead{\colhead{Parameter} & \colhead{QUaD TE/EE/BB} & \colhead{QUaD EE/BB}& \colhead{QUaD TT} & \colhead{QUaD} & \colhead{CMBmany\tablenotemark{a}}} 
    \startdata
    $\Omega_bh^2$ & $0.0233\,\,_{-0.0030}^{+0.0030}$ & $0.0327\,\,_{-0.0098}^{+0.0097}$ & $0.0218\,\,_{-0.0040}^{+0.0040}$& $0.0243\,\,_{-0.0025}^{+0.0025}$&$0.0225\,\,_{-0.0006}^{+0.0006}$\\
    $\Omega_ch^2$ & $0.124\,\,_{-0.030}^{+0.030}$ & $0.162\,\,_{-0.031}^{+0.032}$ &$0.117\,\,_{-0.034}^{+0.036}$ &$0.119\,\,_{-0.0250}^{+0.0253}$ &$0.114\,\,_{-0.006}^{+0.006}$\\
    $\theta$ & $1.040 \pm 0.006$ & $1.032\,\,_{-0.010}^{+0.011}$ &  $1.045\,\,_{-0.011}^{+0.011}$ &$1.041\,\,_{-0.005}^{+0.005}$&$1.041\,\,_{-0.003}^{+0.003}$\\
    $\tau$ \tablenotemark{b} & $ < 0.53 $ ($95$ \% cl) & $ < 0.46 $ ($95$ \% cl) & $ < 0.54 $ ($95$ \% cl)&$ < 0.54 $ ($95$ \% cl) & $0.087\,\,_{-0.017}^{+0.017}$\\
    ln$(10^{10}A_s)$\tablenotemark{c} & $3.53\,\,_{-0.31}^{+0.31}$ &$3.74\,\,_{-0.21}^{+0.19}$& $3.48\,\,_{-0.34}^{+0.34}$& $3.58\,\,_{-0.30}^{+0.28}$ & $3.09\,\,_{-0.04}^{+0.04}$\\
    $n_s$ \tablenotemark{c}& $0.768 \,\,_{-0.151}^{+0.150}$ & $0.530\,\,_{-0.173}^{+0.166}$ & $0.920\,\,_{-0.118}^{+0.117}$  & $0.804\,\,_{-0.098}^{+0.098}$&$0.964\,\,_{-0.013}^{+0.013}$\\
    Age (GYr) & $13.7\,\,_{-0.4}^{+0.4}$ & $13.1\,\,_{-0.93}^{+0.93}$ & $13.6\,\,_{-0.6}^{+0.6}$ & $13.5\,\,_{-0.4}^{+0.4}$&$13.7\,\,_{-0.1}^{+0.1}$\\
    $\Omega_\Lambda$ & $0.64 \pm 0.18$ & $0.45\,\,_{-0.27}^{+0.26}$ &  $0.68\,\,_{-0.20}^{+0.18}$&$0.68\,\,_{-0.14}^{+0.14}$&$0.72\,\,_{-0.03}^{+0.03}$\\
    $\Omega_m$ & $0.36\,\,_{-0.18}^{+0.19}$ & $0.55\,\,_{-0.26}^{+0.27}$ & $0.32\,\,_{-0.18}^{+0.21}$&$0.32\,\,_{-0.14}^{+0.14}$&$0.28\,\,_{-0.03}^{+0.03}$\\
    $z_{re}$ & $23.1\,\,_{-10.0}^{+9.0}$ & $18.4\,\,_{-9.8}^{+9.3}$& $24.3\,\,_{-11.2}^{+10.3}$&$23.9\,\,_{-9.0}^{+8.1}$&$10.5\,\,_{-1.4}^{+1.4}$\\
    $H_0$ & $68.6\,\,_{-12.1}^{+12.5}$ & $63.2\,\,_{-12.1}^{+13.1}$ & $72.5\,\,_{-15.6}^{+15.8}$ &$71.1\,\,_{-10.9}^{+10.9}$&$70.4\,\,_{-2.4}^{+2.4}$\\
    $\sigma_8$ \tablenotemark{d}& $0.98 \pm 0.17$ & $1.07\,\,_{-0.15}^{+0.14}$&  $0.98\,\,_{-0.21}^{+0.21}$&  $0.98\,\,_{-0.15}^{+0.15}$&$0.82\,\,_{-0.03}^{+0.03}$\\
    \enddata
    \tablenotetext{a}{CMBmany describes the CMB data set including ACBAR (up to $l=$2000), CBIPol, BICEP and WMAP5 (see \ref{sec:other}).}
    \tablenotetext{b}{The parameterization of reionization by CAMB has recently been changed. First ionization of helium and hydrogen are now modelled as simultaneous, affecting the returned values of $\tau$ and the time at which $z_{re}$ is defined has changed at the ~6\% level.}
    \tablenotetext{c}{The pivot point for $A_s$ and $n_s$  used with all the data sets presented in this table is $k_p=0.05$~Mpc$^{-1}$.}
    \tablenotetext{d}{The root mean squared linear mass perturbation is defined on a scale of 8$h^{-l}$Mpc, where $H_0=100h$km s$^{-1}$ Mpc$^{-1}$.}
\end{deluxetable*}

\subsection{Parameter Constraints from QUaD data}
\label{sec:qpparams}

We present parameter constraints obtained from QUaD data in Table \ref{tab:paramsQUaD}. The table presents cosmological parameters calculated from the QUaD TE/EE/BB spectra subset, QUaD EE/BB spectra , QUaD TT spectra and the full QUaD data set. The parameters obtained from the subsets are consistent. Cosmological parameter limits from the CMBmany data set are also presented in Table \ref{tab:paramsQUaD}. 

\begin{figure*}[th!]
\resizebox{17.5cm}{!}{\includegraphics{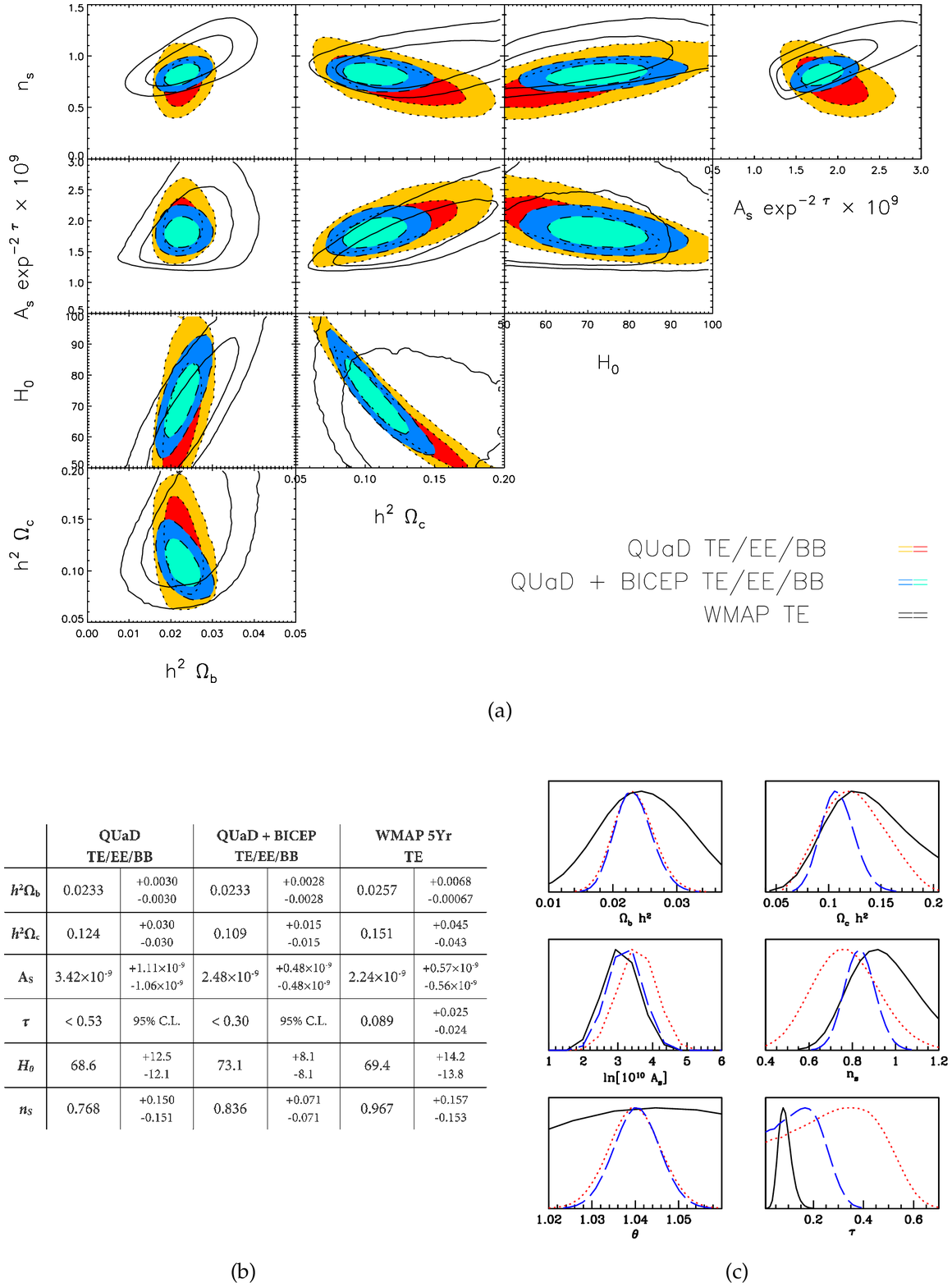}}
\caption{Parameter constraints and contours obtained from QUaD polarization data (TE/ EE/ BB) and QUaD polarization combined with BICEP polarization using a pivot scale of $k_p=0.05$: (a)2D contours at 95\% and 68\% levels for the traditional physical parameter set making use of the combined $A_s\exp{-2}\tau$ distribution; (b) constraints on the six traditional physical parameters; (c)1D marginalized distributions of Monte Carlo Markov chain baseline parameters; QUaD polarization only constraints red/orange dotted; QUaD + BICEP polarization constraints blue/green dashed; constraints from WMAP 5 year polarization data black solid lines.}
\label{fig:polparams}
\end{figure*}

In Figure \ref{fig:polparams} we display the parameter constraints and contours obtained from QUaD polarization data (TE/EE/BB) alone. The resultant confidence regions are considerably smaller than the corresponding confidence regions from WMAP 5 year TE data, over-plotted in Figure \ref{fig:polparams}, for all parameters except $A_s$ and $\tau$. The parameter constraints arrived at using QUaD TE/EE/BB are within 95\% agreement with the parameter constraints from the full WMAP data set and those from WMAP TE spectrum. We have carried out, in addition, a combined analysis using QUaD and BICEP polarization data, these latest-generation ground-based polarization experiments detecting polarization at both low and high-$\ell$. The preferred parameter regions for QUaD and BICEP TE/EE/BB data are also presented in Figure \ref{fig:polparams}. We note that the centre of the preferred range for the scalar spectral index from QUaD TE/EE/BB data is $\sim 0.77$. We also note that this preferred range in $n_s$ concurs with the QUaD polarization analysis of Parameter Paper 1, which found $n_s=0.766 \pm 0.152$. 


\begin{figure}[h!tp]
\resizebox{8cm}{!}{\includegraphics{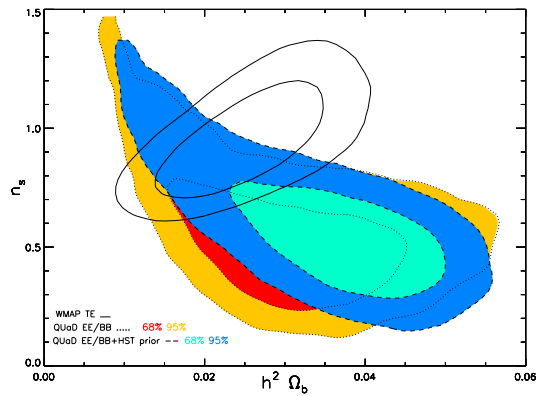}}
\caption{Likelihood contours obtained from QUaD pure polarization data (EE/ BB) at 95\% and 68\% levels for the scalar spectral index and the baryon density, displaying the tail of the distribution to high values of $n_s$. QUaD EE/BB only constraints red/orange dotted; QUaD EE/BB constraints with a Gaussian prior on the Hubble constant blue/green dashed; constraints from WMAP 5 year polarization data black solid lines.}
\label{fig:EB_plot}
\end{figure}

The best-fit spectral index value obtained from the QUaD EE/BB spectra subset is 0.53. We show in Figure \ref{fig:EB_plot} that the likelihood surface is skewed over the range of scalar spectral index values. We over-plot the likelihood for WMAP TE to demonstrate the best-fit regions for the spectral index overlap. To supplement our QUaD EE/BB only parameter fits we also carried out an analysis where we implement a Gaussian prior on the Hubble constant with the latest independent limits obtained by the Hubble Space telescope \citep[HST,][]{HST2} $74.2\pm3.6$kms$^{-1}$Mpc$^{-1}$. The Hubble constant primarily impacts the width and separation of the spectral peaks, while the spectral index is evaluated from the {\it tilt} of the data. Applying the information we have on the Hubble constant improves our parameter confidence ranges, while not preferentially impacting the spectral index signal in the data. These contours are also plotted in Figure \ref{fig:EB_plot}, confirming the robustness of the EE/BB analysis. The parameter constraints from QUaD EE/BB spectra improve on the corresponding constraints we presented in Parameter Paper 1.

The constraint obtained on the acoustic peak scale, $\theta$, is as tight from QUaD EE/BB only as it is from the QUaD temperature spectrum and the constraint on $h^2\Omega_{c}$ is tighter.

\begin{figure*}[ht]
\resizebox{17cm}{!}{\includegraphics{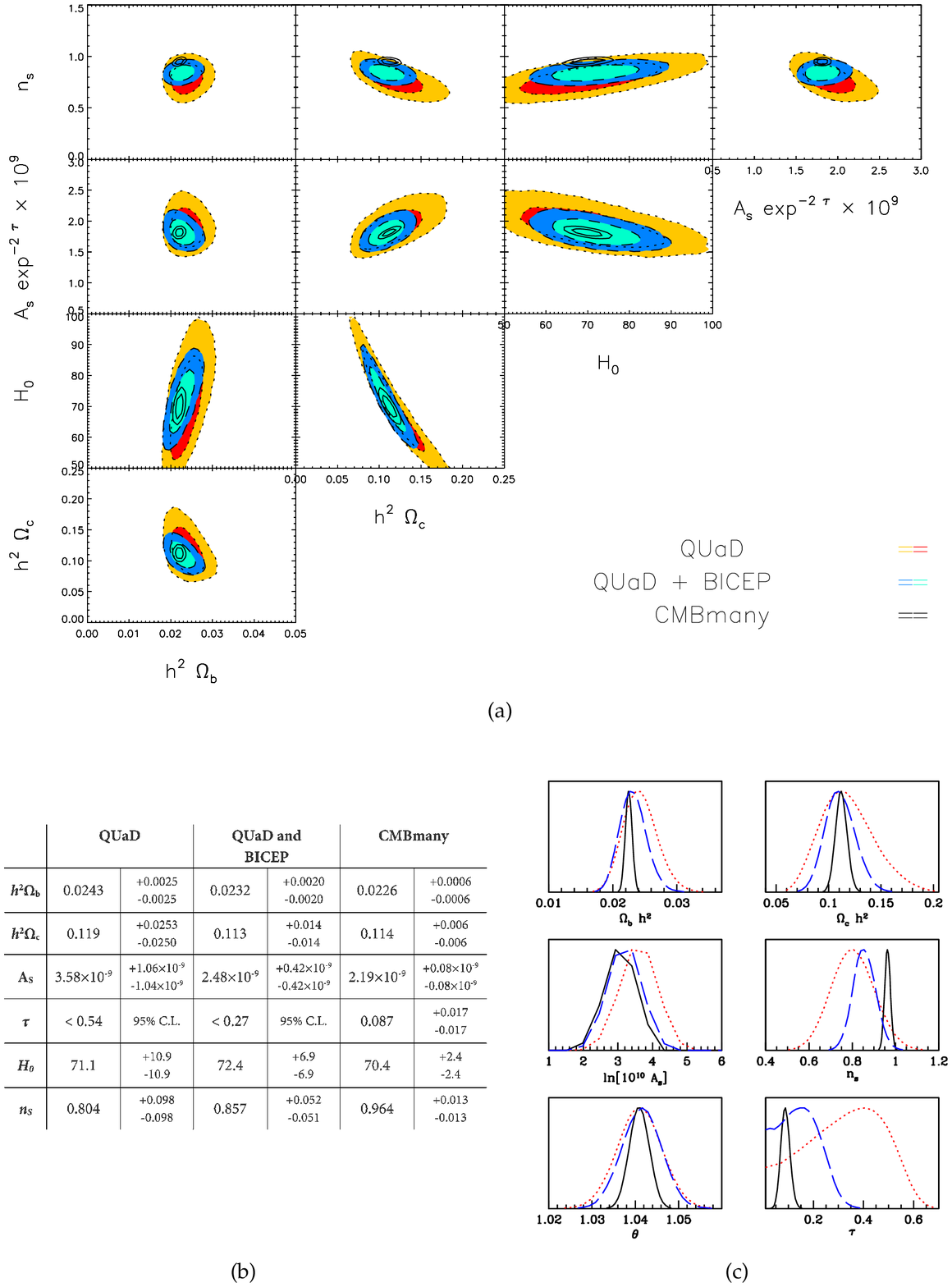}}
\caption{Parameter constraints and contours obtained from QUaD temperature and polarization data using a pivot scale of $k_p=0.05$: (a)2D contours at 95\% and 68\% levels for the traditional physical parameter set making use of the combined $A_s\exp{-2}\tau$ distribution; (b) constraints on the six traditional physical parameters; (c)1D marginalized distributions of Monte Carlo Markov chain baseline parameters; QUaD temperature and polarization constraints red/orange dotted; QUaD + BICEP temperature and polarization constraints blue/green dashed; constraints from CMBmany: (ACBAR/ WMAP/ CBI/ BICEP) (see \ref{sec:other}) black solid lines. The best-fit value of $\tau$ from QUaD temperature and polarization data is 0.314.}
\label{fig:2D_Q08_main}
\end{figure*}

The parameter ranges arrived at by using the complete QUaD data set agree with the values from the CMBmany data set within the 95\% contour limits, as demonstrated in Figure \ref{fig:2D_Q08_main}. The preferred parameter range for the baryon density we presented in Parameter Paper 1 from the full \cite{pryke09} QUaD data set was $\sim 50\%$ higher than that we now provide. The Hubble constant was also $\sim 30\%$ higher than the value we return from the analysis of the new non-field-differenced data. The data analysis pipeline which produced the \cite{Michael09} spectra differs from the \cite{pryke09} analysis in the implementation of a new noise-removal strategy and in modelling a new beam shape. This results in a material difference in the spectra in addition to a reduction of cosmic variance. The scalar spectral amplitude of Parameter Paper 1 (which we quoted as part of a combined $A_s\exp{(-2\tau)}$ parameter) was significantly shifted relative to our current preferred range, the corollary being in extended parameter analysis isocurvature ratios were more constrained. However, as our parameter analysis has evolved the confidence limits presented in Parameter Paper 1 cannot be directly compared to the limits in this paper. Repeating the parameter analysis of the second and third season QUaD data of \cite{pryke09}, using our current CosmoMC-based pipeline, we find 68\% confidence regions for $A_s\exp{(-2\tau)}$ and the Hubble constant that are 30\% greater than the corresponding values obtained using the improved data set of \cite{Michael09}.

\subsection{Parameter Constraints Combining Data Sets.}
\label{sec:comb}

We continue our parameter analysis by investigating the effect of adding QUaD data to the CMBmany temperature and polarization  ensemble data set. 

\begin{figure}
\resizebox{\columnwidth}{!}{\includegraphics{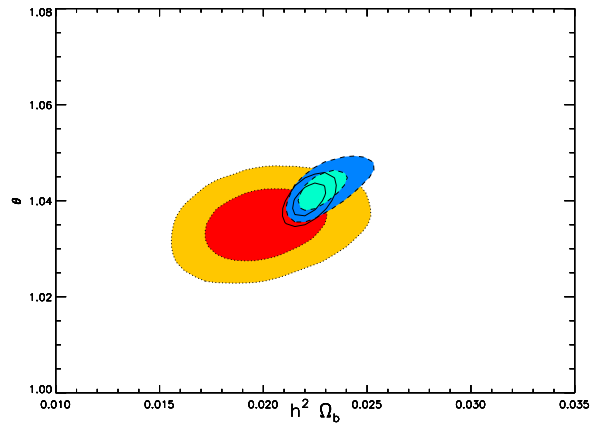}}
\caption{2D marginalized contours $h^2 \Omega_b$ against $\theta$ from separate temperature and polarization combined QUaD and CMBmany data sets. QUaD polarization + CMBmany polarization constraints red/orange dotted; QUaD temperature + CMBmany temperature constraints blue/green dashed; constraints from QUaD + CMBmany: black solid lines.}
\label{fig:Theta_TallPall}
\end{figure}

In order to investigate the influence of polarization in our cosmological analysis, we carried out cosmological parameter fits using the TT and TE/EE/BB subsets of the QUaD+CMBmany data set. The resulting best-fit regions, presented in Table \ref{tab:paramscomb2} and  Figure \ref{fig:Theta_TallPall}, are consistent, although the polarization spectra prefer lower values for the Hubble constant and acoustic peak scale as well as a lower spectral index. The multi-experiment TE/EE/BB data set contributes information on all six cosmological parameters.
\vspace{1cm}

\begin{deluxetable*}{l c c c c}[p!]
  \tablewidth{12cm}
  \tablecaption{
	Cosmological Parameter Constraints using combined CMB Temperature and Polarization   
\label{tab:paramscomb2}}
\tablehead{\colhead{Parameter} & \colhead{QUaD + CMBmany TE/EE/BB\tablenotemark{a}} & \colhead{QUaD + CMBmany TT\tablenotemark{b}}& \colhead{QUaD + CMBmany}} 
    \startdata
    $\Omega_bh^2$ & $0.0204 \pm 0.0020$  & $0.0231 \pm 0.0008$& $0.0224 \pm 0.0005$\\
    $\Omega_ch^2$ & $0.124 \pm 0.014$ & $0.109\,\,_{-0.007}^{+0.007}$& $0.113\,\,_{-0.005}^{+0.005}$\\
    $\theta$ & $1.036 \pm 0.005$ &$1.042\,\,_{-0.002}^{+0.003}$&$1.041\,\,_{-0.002}^{+0.003}$\\
    $\tau$ \tablenotemark{c} &$0.073\,\,_{-0.017}^{+0.017}$&$0.127\,\,_{-0.079}^{+0.079}$&$0.086\,\,_{-0.017}^{+0.017}$\\
    ln$(10^{10}A_s)$ \tablenotemark{d}& $3.22\,\,_{-0.08}^{+0.08}$ & $3.17\,\,_{-0.15}^{+0.15}$ & $3.13\,\,_{-0.03}^{+0.04}$\\
    $n_s$ \tablenotemark{d}& $0.842 \pm 0.055$& $0.983\,\,_{-0.025}^{+0.026}$& $0.960\,\,_{-0.013}^{+0.013}$\\
    Age (GYr) & $14.1\,\,_{-0.3}^{+0.3}$ & $13.6\,\,_{-0.2}^{+0.2}$ & $13.6\,\,_{-0.2}^{+0.2}$  \\
    $\Omega_\Lambda$ & $0.63 \pm 0.10$& $0.75\,\,_{-0.04}^{+0.04}$ & $0.73\,\,_{-0.03}^{+0.03}$ \\
    $\Omega_m$ & $0.37 \pm 0.10$ & $0.25\,\,_{-0.04}^{+0.04}$ & $0.27\,\,_{-0.03}^{+0.03}$ \\
    $z_{re}$ & $10.1\,\,_{-1.5}^{+1.5}$ &$12.6\,\,_{-5.7}^{+5.5}$ &$10.4\,\,_{-1.4}^{+1.4}$ \\
    $H_0$ & $63.8\,\,_{-6.1}^{+6.1}$& $73.7\,\,_{-3.8}^{+3.8}$& $70.6\,\,_{-2.3}^{+2.3}$\\
    $\sigma_8$& $0.79\,\,_{-0.07}^{+0.06}$&$0.83\,\,_{-0.05}^{+0.05}$&$0.82\,\,_{-0.03}^{+0.03}$\\
    \enddata
    \tablenotetext{a}{CMBmany TE/EE/BB is the combination of WMAP, CBI and BICEP polarization spectra.}
    \tablenotetext{b}{CMBmany TT is the combination of WMAP, CBI, ACBAR and BICEP temperature spectra.}
    \tablenotetext{c}{The parameterization of reionization by CAMB has recently been changed. First ionization of helium and hydrogen are now modelled as simultaneous, affecting the returned values of $\tau$ and the time at which $z_{re}$ is defined has changed at the ~6\% level.}
    \tablenotetext{d}{The pivot point for $A_s$ and $n_s$ is $k_p=0.013$~Mpc$^{-1}$.}
  \end{deluxetable*}

\section{Conclusions}
\label{sec:concl}

We present best-fit regions for the standard $\Lambda$CDM six-parameter model using QUaD CMB spectra alone, and QUaD data in combination with CMBmany (BICEP, ACBAR, CBI and WMAP).

QUaD data provide an unprecedented amount of independent information over four spectra in a multipole range of $200 < \ell < 2000$. This improves the ensemble of CMB data sets specifically providing independent cohesive information over an angular range broad enough to span the sparsely populated ``hinge'' multipole regions of the CMBmany data set.

Data from the QUaD experiment, as a single, discerning data set can provide excellent independent measures of the baryon density in the universe, the cold dark matter density, and the acoustic scale, which measures the universe's evolution history.

QUaD TE/EE/BB spectra provide considerably tighter constraints on four of the six standard $\Lambda$CDM parameters than can be obtained from 5 year WMAP polarization data. We present the best confidence ranges obtained from QUaD EE/BB spectra alone. The QUaD temperature spectrum alone also provides good parameter constraints in excellent agreement with the WMAP 5 year best-fit parameters. 


\acknowledgements QUaD is funded by the National Science Foundation in
the USA, through grants ANT-0338138, ANT-0338335 \& ANT-0338238, by
the Science and Technology Facilities Council (STFC) in the UK and by
the Science Foundation Ireland. PGC acknowledges funding from the 
Portuguese FCT. SEC acknowledges support
from a Stanford Terman Fellowship. JRH acknowledges the support of an
NSF Graduate Research Fellowship, a Stanford Graduate Fellowship and a
NASA Postdoctoral Fellowship.  YM acknowledges support from a SUPA
Prize studentship. CP acknowledges partial support from the Kavli
Institute for Cosmological Physics through the grant NSF PHY-0114422.
EYW acknowledges receipt of an NDSEG fellowship. MZ acknowledges
support from a NASA Postdoctoral Fellowship. We acknowledge
the use of the  CAMB \citep{lewis00}, CosmoMC
\citep{lewis02} and HEALPix \citep{gorski05} packages. We acknowledge
the use of the Legacy Archive for Microwave Background Data Analysis
(LAMBDA). Support for LAMBDA is provided by the NASA Office of Space
Science. This work was performed using the computational facilities of 
the Advanced Research Computing @ Cardiff (ARCCA) Division, Cardiff 
University.

\bibliographystyle{apj}

\bibliography{cosmo}

\begin{thebibliography}{}

\bibitem[\protect\citeauthoryear{{Ade} et~al.}{{Ade} et~al.}{2008}]{ade07}
{Ade}, P., et~al. 2008, \apj, 674, 22

\bibitem[\protect\citeauthoryear{{Bond}, {Jaffe}, \& {Knox}}{{Bond}
  et~al.}{2000}]{BJK2000}
{Bond}, J.~R., {Jaffe}, A.~H.,  \& {Knox}, L. 2000, \apj, 533, 19

\bibitem[\protect\citeauthoryear{{Bridle} et~al.}{{Bridle}
  et~al.}{2002}]{bridle02}
{Bridle}, S.~L., {Crittenden}, R., {Melchiorri}, A., {Hobson}, M.~P.,
  {Kneissl}, R.,  \& {Lasenby}, A.~N. 2002, \mnras, 335, 1193

\bibitem[\protect\citeauthoryear{{Brown} et~al.}{{Brown}
  et~al.}{2009}]{Michael09}
{Brown}, M.~L., et~al. 2009, \apj, 705, 978

\bibitem[\protect\citeauthoryear{{Castro} et~al.}{{Castro}
  et~al.}{2009}]{Castro09}
{Castro}, P.~G., et~al. 2009, \apj, 701, 857

\bibitem[\protect\citeauthoryear{{Chiang} et~al.}{{Chiang}
  et~al.}{2010}]{BICEP}
{Chiang}, H.~C., et~al. 2010, \apj, 711, 1123

\bibitem[\protect\citeauthoryear{{Dunkley} et~al.}{{Dunkley}
  et~al.}{2009}]{dunkleypol}
{Dunkley}, J., et~al. 2009, \apj, 701, 1804

\bibitem[\protect\citeauthoryear{{Friedman} et~al.}{{Friedman}
  et~al.}{2009}]{Friedman09}
{Friedman}, R.~B., et~al. 2009, \apjl, 700, L187

\bibitem[\protect\citeauthoryear{{Gelman} \& {Rubin}}{{Gelman} \&
  {Rubin}}{1992}]{gelman92}
{Gelman}, A.,  \& {Rubin}, D.~B. 1992, Statist. Sci., 7, 457

\bibitem[\protect\citeauthoryear{{G{\'o}rski} et~al.}{{G{\'o}rski}
  et~al.}{2005}]{gorski05}
{G{\'o}rski}, K.~M., {Hivon}, E., {Banday}, A.~J., {Wandelt}, B.~D., {Hansen},
  F.~K., {Reinecke}, M.,  \& {Bartelmann}, M. 2005, \apj, 622, 759

\bibitem[\protect\citeauthoryear{{Hedman} et~al.}{{Hedman}
  et~al.}{2002}]{Hedman}
{Hedman}, M.~M., {Barkats}, D., {Gundersen}, J.~O., {McMahon}, J.~J., {Staggs},
  S.~T.,  \& {Winstein}, B. 2002, \apjl, 573, L73

\bibitem[\protect\citeauthoryear{{Hinderks} et~al.}{{Hinderks}
  et~al.}{2009}]{Hinderks09}
{Hinderks}, J.~R., et~al. 2009, \apj, 692, 1221

\bibitem[\protect\citeauthoryear{{Jones} et~al.}{{Jones}
  et~al.}{2006}]{jones06}
{Jones}, W.~C., et~al. 2006, \apj, 647, 823

\bibitem[\protect\citeauthoryear{{Komatsu} \& {Seljak}}{{Komatsu} \&
  {Seljak}}{2002}]{SZ}
{Komatsu}, E.,  \& {Seljak}, U. 2002, \mnras, 336, 1256

\bibitem[\protect\citeauthoryear{{Kovac} et~al.}{{Kovac}
  et~al.}{2002}]{kovac02}
{Kovac}, J.~M., {Leitch}, E.~M., {Pryke}, C., {Carlstrom}, J.~E., {Halverson},
  N.~W.,  \& {Holzapfel}, W.~L. 2002, Nature, 420, 772

\bibitem[\protect\citeauthoryear{{Lewis} \& {Bridle}}{{Lewis} \&
  {Bridle}}{2002}]{lewis02}
{Lewis}, A.,  \& {Bridle}, S. 2002, \prd, 66, 103511

\bibitem[\protect\citeauthoryear{{Lewis}, {Challinor}, \& {Lasenby}}{{Lewis}
  et~al.}{2000}]{lewis00}
{Lewis}, A., {Challinor}, A.,  \& {Lasenby}, A. 2000, \apj, 538, 473

\bibitem[\protect\citeauthoryear{{Nolta} et~al.}{{Nolta}
  et~al.}{2009}]{nolta08}
{Nolta}, M.~R., et~al. 2009, \apjs, 180, 296

\bibitem[\protect\citeauthoryear{{O'Sullivan} et~al.}{{O'Sullivan}
  et~al.}{2008}]{osullivan08}
{O'Sullivan}, C., et~al. 2008, Infrared Physics and Technology, 51, 277

\bibitem[\protect\citeauthoryear{{Pryke} et~al.}{{Pryke}
  et~al.}{2009}]{pryke09}
{Pryke}, C., et~al. 2009, \apj, 692, 1247

\bibitem[\protect\citeauthoryear{{Readhead} et~al.}{{Readhead}
  et~al.}{2004}]{readhead04}
{Readhead}, A.~C.~S., et~al. 2004, Science, 306, 836

\bibitem[\protect\citeauthoryear{{Reichardt} et~al.}{{Reichardt}
  et~al.}{2009}]{reichardt08}
{Reichardt}, C.~L., et~al. 2009, \apj, 694, 1200

\bibitem[\protect\citeauthoryear{{Riess} et~al.}{{Riess} et~al.}{2009}]{HST2}
{Riess}, A.~G., et~al. 2009, \apj, 699, 539

\bibitem[\protect\citeauthoryear{{Sievers} et~al.}{{Sievers}
  et~al.}{2009}]{Sievers09}
{Sievers}, J.~L., et~al. 2009, arXiv e-prints 0901.4540

\end{thebibliography}

\end{document}